\documentclass[prd,showpacs,nofootinbib,
12pt]{revtex4}

\usepackage[utf8]{inputenc}
\usepackage[russian]{babel}
\usepackage{amssymb,latexsym,amsmath,amscd}
 \usepackage[pdftex]{graphicx}

\begin{document}

\newcommand{\Pom}{{\hspace{-0.1em}I\hspace{-0.25em}P}}

\title{Классические глюонные поля релятивистских цветовых зарядов}
\author{\firstname{А. С.}~\surname{Задора}}
\email{zadorra@yandex.ru}
\affiliation {Московский государственный университет им. М.В. Ломоносова, 
физический факультет, 
Москва, Россия}

\begin{abstract}
\noindent
Целью настоящей работы является более детальное рассмотрение недавно обнаруженного необычного эффекта сияния цветового заряда и его возможных физических проявлений, связанных с рассмотрением ансамблей частиц с цветовым зарядом на классическом уровне. Исследуются способы появления данного эффекта в произвольных системах точечных массивных частиц, которые могут быть разбиты на элементарные конфигурации. Также рассматривается возможное влияние этого эффекта на динамику частиц (в частности, на функцию распределения глюонов). Для естественной регуляризации выражений рассматриваются столкновения с заданным прицельным параметром. Показано, что в случае разумных значений прицельного параметра столкновения  могут проходить в ``электродинамическом'' режиме, при этом вклад ``сияния'' зарядов в напряженности полей оказывается подавленным по сравнению с электродинамической картиной. Исходя из анализа ситуации с ``цветовым эхо'' следует, что сделанный вывод оказывается справедливым и для более сложных конфигураций частиц, так как жесткие глюонные поля могут генерироваться только в результате непосредственного столкновения, а не за счет каких-либо эффектов типа ``эха''.

\noindent
\end{abstract}

\maketitle

\newpage
\section{ВВЕДЕНИЕ}
Классические решения неабелевых полевых уравнений и их свойства неоднократно вызывали интерес в связи с их нелинейным характером в отличие от уравнений классической электродинамики. Немаловажную роль в этом сыграло и явление конфайнмента кварков. Однако, в отличие от электродинамики, напряженности неабелевых полей не являются непосредственными наблюдаемыми величинами, и более адекватным становится описание на языке квантовой теории (КХД). Классические поля можно трактовать как средние значения соответствующих операторов по некоторому состоянию в пределе больших квантовых чисел. \\

В рамках другого подхода, связанного с разложением по константе связи, справедливого при высоких энергиях, рассматриваются вопросы, касающиеся неабелевого излучения, например, аналога синхротронного в электродинамике 
~\cite{tuchin, dbeyssi}.  \\

В последнее время интерес к классическим конфигурациям полей возрос в связи с необходимостью использования транспортных кодов (где эти 
конфигурации могут быть включены, см., например, модификацию транспортного кода HSD ~\cite{toneev}, которая учитывает электромагнитные поля на классическом уровне) для описания
динамики систем в релятивистских столкновениях тяжелых ионов. В задачах релятивистских столкновений широкой популярностью пользуется модель так называемого конденсата цветового стекла (CGC), в которой квазиклассические поля являются источником глюонных конфигураций в переменных светового конуса (см., например, ~\cite{mclerran} или обзор ~\cite{muller}). \\

Недавно было обнаружено ~\cite{dipodip}, что при лобовых столкновениях ультрарелятивистских частиц с открытым цветом глюонные конфигурации напоминают электродинамические, однако сигнал от момента встречи существенно отличается. Именно, область наблюдения порядка нескольких ферми на короткое время оказывается заполнена глюонными полями огромной амплитуды (так называемое сияние цветового заряда), напоминающими ударную волну. Этот эффект не улавливается, например, в модели CGC в связи с тем, что работа ведется на световом конусе. Соответствующие полевые конфигурации могут быть важны при анализе процессов, происходящих на ранней стадии формирования кварк-глюонной плазмы, когда рождается большое количество кварк-антикварковых пар, моделью которых может выступать цветовой диполь. В ~\cite{dipoles} исследовались поля Вайцзеккера -- Вильямса в квазиклассическом приближении для ультрарелятивистской системы  ``частица-диполь'', после чего результат сравнивался с аналогичным из CGC. Было показано, что в случае ориентированного вдоль направления движения диполя предел скоростей, близких к скорости света ($v \to 1$), существует для усредненных по времени столкновения с диполем полевых конфигураций. При этом усредненные значения напряженностей полей оказываются несингулярными, и функция распределения качественно совпадает с результатом работы ~\cite{struct}. Однако в ~\cite{dipoles} не учитывался эффект сияния цветового заряда, так как он появляется в следующем порядке разложения по константе связи. \\

В ~\cite{dipodip} исследовались напряженности глюонных полей для трех простейших базовых конфигураций: лобовые столкновения систем типа ``частица-частица'', ``частица-диполь'', ``диполь-диполь''. Регуляризация выражений (в силу отсутствия прицельного расстояния) производилась путем обрезания решений на искусственном временном масштабе $t _{min}$ и последующей сшивкой.  Обнаруживается так называемое ``сияние цветового заряда'', обусловленное вращением изовекторов зарядов в изотопическом пространстве, которое усиливается при переходе к более сложным конфигурациям. Сигнал от этого эффекта длится достаточно малые промежутки времени порядка $10^{-4}$ Фс, но охватывает всю область наблюдения порядка нескольких ферми. В силу приближенности решения не ухватывается эффект ``цветового эха'', который заключается в приходе к третьей частице сигнала о вращении с момента встречи двух частиц.\\

Дабы несколько прояснить ситуацию, будем рассматривать задачу, аналогичную поставленной в ~\cite{dipodip}, но с заданным прицельным параметром, выступающим в качестве естественного регуляризатора выражений. 

\section{Поле классического заряда в неабелевой теории}
Для начала напомним, как выглядит ситуация в абелевой теории - электродинамике. \\
В электродинамике поле точечного заряда $e$, движущегося по траектории $\mathbf{r}(t)$ со скоростью $\mathbf{v}(t)$, выражается при помощи потенциалов Лиенара -- Вихерта ~\cite{landau}
\begin{equation} \label{Lienar}
\varphi = \frac{1}{4\pi}\left [\frac{e}{R - \mathbf{v}\mathbf{R}} \right ]_{t'} , \ \ \mathbf{A} = \frac{1}{4\pi} \left [ \frac{e\mathbf{v}}{R - \mathbf{vR}} \right ]_{t'} \ .
\end{equation}
Запаздывающее время $t'$ определяется из уравнения 
\begin{equation}\label{delay}
R(t') = t-t' \  ,
\end{equation}
где $R = | \mathbf{R} |$ -- расстояние от точки наблюдения до частицы. \\
Напряженности электрического и магнитного полей даются выражениями:
\begin{equation} \label{EH}
\mathbf{E} = -\frac{\partial \mathbf{A}}{\partial t} - \mathbf{\nabla} \varphi , \ \ \mathbf{H} = \mathbf{\nabla} \times \mathbf{A}.
\end{equation}
Из (\ref{EH}) с помощью (\ref{Lienar}), (\ref{delay}) получают:
\begin{gather}
\mathbf{E} = \frac{1}{4\pi} \left [ \frac{e}{R^2} \frac{(1-v^2) (\mathbf{n} - \mathbf{v})}{(1-\mathbf{v}\mathbf{n})^3} + \frac{e}{R} \frac{\mathbf{n} \times (\mathbf{n} - \mathbf{v}) \times \dot {\mathbf{v}} }{(1-\mathbf{v}\mathbf{n})^3} \right ]_{t'} \ ,\\
\mathbf{H} = \mathbf{n} \times \mathbf{E}. \nonumber
\end{gather}

\hspace{\parindent}

Естественным образом встает вопрос о возможности подобного описания систем из релятивиcтских цветовых зарядов, т. е. использования аналогов потенциалов Лиенара -- Вихерта в неабелевой теории. При этом обнадеживает то, что для одной частицы во всех порядках по константе взаимодействия $g$ из уравнений Янга -- Миллса получается кулоновское решение. Для системы из двух частиц в нулевом приближении по $g$ имеют место  аналогичные кулоновские решения. Для интересующих релятивистских систем эффективная константа связи $\alpha_g \sim 0.4$, что позволяет строить ряды теории возмущений, используя только несколько первых порядков. 

Для теории с лагранжианом (для простоты рассматриваем группу $SU(2)$)
\begin{equation}
\mathcal{L} = -\frac{1}{4} \widetilde{{G_{\mu \nu}}} \widetilde{{G^{\mu \nu}}} - \widetilde{{j^\mu}} \widetilde{{A_\mu}}  \ ,
\end{equation}
где $\widetilde{A}_\mu = \left (A_\mu ^ 1, A_\mu ^2, A_\mu ^3  \right )$, тензор поля $\widetilde{{G_{\mu \nu}}} = \partial _\mu \widetilde{A_\nu} - \partial _\nu \widetilde{A_\mu} + g \widetilde{A_\mu} \times \widetilde{A_\nu}$, ковариантная производная действует по правилу $\widetilde{D^\mu} \widetilde{f} =  \partial ^\mu \widetilde{f} + g \widetilde{A^\mu} \times \widetilde{f}$. Уравнения поля
\begin{equation}
\widetilde {D^\mu} \widetilde{G_{\mu \nu}} = \widetilde{j_\nu}
\end{equation}
для системы из двух цветовых зарядов $\widetilde{P}, \widetilde{Q}$ факторизуются, и удается получить так называемые уравнения хромостатики ~\cite{khriplovich}
\begin{gather}\label{chromostat}
\mathbf{D}\mathbf{D} \Phi = \delta, \\
\nabla \times \nabla \times \mathbf{a} = g \mathbf{j} \ ,  \mathbf{j} = \Phi J \mathbf{D} \Phi. \nonumber
\end{gather} 
Здесь $\Phi = \varphi - \star{\varphi}$ -- двухкомпонентный столбец, где $\varphi_1 = \varphi_1 (\mathbf{x}), \varphi_2 = \varphi_2(\mathbf{x}), \star{\varphi_1} = \varphi_1(\mathbf{x_2}), \star{\varphi_2} = \varphi_2(\mathbf{x_1})$, $\delta =  || \delta (\mathbf{x} - \mathbf{x_1}) \ , \  \delta (\mathbf{x} - \mathbf{x_2}) || $, ковариантная производная определена как $\mathbf{D}_{kl} = \nabla \delta_{kl} + g \mathbf{a} C_{kl}, k, l = 1,2$, а $C$ и $J$ есть матрицы:
\begin{equation}
C = \begin{pmatrix}
 -(\widetilde{P} \widetilde{Q}) \ \ \ -(\widetilde{Q} \widetilde{Q}) \\
(\widetilde{P} \widetilde{P}) \ \ \ (\widetilde{P} \widetilde{Q}) 
\end{pmatrix},  \ \ J = \begin{pmatrix}
0 \ \ 1 \\
-1 \ \ 0
\end{pmatrix}.
\end{equation}
При этом векторное поле $\mathbf{a}$ оказывается натянутым на вектор $\widetilde{P} \times \widetilde{Q}$ ($\widetilde{P}, \widetilde{Q}$ -- изовекторы цветовых зарядов частиц), а скалярное $\phi$ -- на векторы $\widetilde{P}$ и $\widetilde{Q}$. 
Физический смысл системы уравнений (\ref{chromostat}) состоит в том, что порожденное системой из двух зарядов глюонное поле само является источником заряда. Более того, оно также является и источником ``тока''.
Из данной системы уравнений можно извлечь, что векторное поле $\mathbf{a}$, генерируемое скалярной компонентой $\Phi$, находится в следующем порядке по константе связи по сравнению с $\Phi$. Система уравнений (\ref{chromostat}) была исследована как аналитически, так и численно ~\cite{2body}, и было показано, что для интересующих констант связи $g^2/4\pi < \sqrt{2}$ решение хорошо аппроксимируется кулоновскими потенциалами $\varphi$, при этом векторное поле напоминает поле постоянного магнита. \\

Первое уравнение (\ref{chromostat}) можно преобразовать к виду:
\begin{equation}
\mathbf{}\nabla \mathbf{E} = \delta - g \mathbf{a} C \mathbf{E},
\end{equation}
где столбец хромоэлектрического поля определяется как $\mathbf{E} = \mathbf{D} \Phi$. Дополнительное к $\delta$ слагаемое в правой части можно трактовать как плотность заряда порожденного глюонного поля 
\begin{equation}
G = -g \mathbf{a} C \mathbf {E}.
\end{equation}
Так как $\widetilde{P} = -\widetilde{Q}$, то $G_1 = G_2 = \mathbf{a} \mathbf{\nabla} (\Phi_2 - \Phi_1)$, $\widetilde{G} = G_1 \widetilde{P} + G_2 \widetilde{Q} = 0$. Следовательно диполь можно рассматривать как электродинамический. Даже в случае параллельных зарядов $\widetilde{P} = \widetilde{Q}$ получим, что $G_1 = -G_2 = \mathbf{a} \mathbf{\nabla} (\Phi_1 + \Phi_2)$, откуда снова получим $\widetilde{G} = 0$. В случае иных конфигураций зарядов выражение для $G$ является нетривиальным и требует аккуратного учета.\\

В связи с рассмотренным выше в ~\cite{dipodip} предлагается использовать приближенное решение уравнений Янга -- Миллса для двух и более частиц в виде суперпозиции ``потенциалов Лиенара-Вихерта'' для каждой частицы:
\begin{equation}
\varphi = \frac{1}{4\pi}\left [\frac{\widetilde{C}}{R - \mathbf{v}\mathbf{R}} \right ]_{t'}, \ \ \mathbf{A} = \frac{1}{4\pi} \left [ \frac{\widetilde{C}\mathbf{v}}{R - \mathbf{vR}} \right ]_{t'},
\end{equation}
$\widetilde{C}$ -- изовектор цветового заряда частицы.
В результате с помощью стандартной процедуры получаются выражения для хромоэлектрического и хромомагнитного полей движущейся частицы:
\begin{gather}\label{EHNA}
\widetilde{\mathbf{E}} = \frac{1}{4\pi} \left [ \frac{\widetilde{C}}{R^2} \frac{(1-v^2) (\mathbf{n} - \mathbf{v})}{(1-\mathbf{v}\mathbf{n})^3} + \frac{\widetilde{C}}{R} \frac{\mathbf{n} \times (\mathbf{n} - \mathbf{v}) \times \dot {\mathbf{v}} }{(1-\mathbf{v}\mathbf{n})^3}  + \frac{\widetilde{D}}{R} \frac{ (\mathbf{n} - \mathbf{v})}{(1-\mathbf{v}\mathbf{n})^2}\right ]_{t'}, \\
\widetilde{\mathbf{H}} = \mathbf{n} \times \widetilde{\mathbf{E}}. \nonumber
\end{gather}
В отличие от электродинамики здесь появляется дополнительный член следующего порядка по $g$, связанный с изменением направления изовектора заряда ${\widetilde{D}} = \partial \widetilde{C} / \partial t$ (вращения в изотопическом пространстве), следующего из закона сохранения тока (уравнений совместности):
\begin{equation} \label{condition}
\widetilde {D^\mu} \widetilde{j_\mu} = 0
\end{equation}
(в отличие от электродинамики $\partial \mu j_\mu \neq 0$). \\

В ~\cite{dipodip} были получены полуаналитические формулы для вращений изовекторов цветовых зарядов для простейших конфигураций, на которые, как представляется, возможно разложить любой сложный процесс столкновения тяжелых ионов,  типа ``частица-частица'', ``частица-диполь'' и ``диполь-диполь''. Полученные графики напряженностей глюонных полей, генерируемых при лобовом столкновении данных конфигураций в целом проявляют электродинамические черты, однако в момент встречи картина существенно отличается от электродинамической -- генерируемые поля имеют гигантскую амплитуду (так называемое ``сияние цветового заряда'') вследствие формально бесконечно быстрого вращения изовекторов цветовых зарядов при нулевом расстоянии между точечными частицами в момент столкновения (для устранения бесконечности точка момента встречи окружалась окрестностью, на границах которой происходила сшивка решений). Также был упомянут эффект, который не учитывался при построении графиков -- эффект ``цветового эха'', являющийся проявлением коллективной динамики цветовых зарядов при наличии более чем двух частиц. \\

Будем решать задачу, аналогичную поставленной в ~\cite{dipodip}, но с учетом прицельного расстояния. Обозначим прицельное расстояние как $a$. Частицы будем считать точечными и массивными. Исследование неточечных объектов весьма затруднительно, так как в этом случае возникает рассогласование вращения заряда на самой частице. 

\section{Поле двух цветовых зарядов}
Аналогично работе ~\cite{dipodip} будем считать, что взаимодействие ''начинается'' с некоторого момента, когда частицы сближаются на расстояние $D \sim 1$ ферми. Такая постановка задачи является оправданной, поскольку на таких масштабах и при релятивистских скоростях константа связи невелика, что позволяет рассматривать задачу с помощью потенциалов Лиенара -- Вихерта. Траектории частиц считаются фиксированными (так называемое эйкональное приближение), ускорением зарядов пренебрегается. В этом случае естественно рассматривать прицельные параметры, удовлетворяющие условию $a<D$. Выберем следующую лабораторную систему координат: положения  частиц задаются как $x_1 = 0$, $x_2 = a$, $z_1 = v t$, $z_2 = -w t$, $v, w > 0$, а время отсчитывается таким образом, что в момент $t=0$ частицы сближаются в указанной системе координат на минимальное расстояние, равное $a$ (указанный момент сближения на минимальное расстояние далее для краткости условно будем называть моментом столкновения). Обозначим время сближения частиц на расстояние $D$ через $T$. Очевидно, выполняется соотношение: 
\begin{equation}
T = -\frac{\sqrt{D^2 - a^2}}{(v+w)}.
\end{equation}
Далее, если не оговорено иначе, будем рассматривать два значения прицельного параметра $a$: $a = 0.5$, $a = 0.05$ (в единицах ферми). Соответственно, время будем измерять в единицах Фс. 
Далее рассмотрим время прихода сигнала о сближении на расстояние $D$ к частицам. Решая уравнение (\ref{delay}) с нужными значениями $t'$ (в данном случае полагая $t'$ = T), можно получить: 
\begin{gather}
t_1' = \frac{T + Tvw + \sqrt{T^2(v+w)^2 + a^2(1-v^2)}}{1-v^2}, \\
t_2' = \frac{T + Tvw + \sqrt{T^2(v+w)^2 + a^2(1-w^2)}}{1-w^2}. \nonumber
\end{gather}
$t_1'$, $t_2'$ -- времена прихода сигнала о сближении на расстояние $D$ к первой и второй частицам соответственно. 
С этого момента частицы начинают взаимодействовать, т.е. векторы их зарядов начинают вращаться согласно уравненияю (\ref{condition}). Далее нужно рассмотреть время прихода сигнала о начале вращения к частицам, так как это изменит вид уравнений. Обозначим соответствующий момент времени как $t''$ (он одинаков для обеих частиц):
\begin{equation}
t^{''} = \frac{t_2' + t_2'vw + \sqrt{t_2'^2(v+w)^2 + a^2(1-v^2)}}{1-v^2}.
\end{equation}
Таким образом, задача разбивается на три этапа (см. далее), что позволяет строить решение на каждом из этапов в отдельности, должным образом сшивая затем полученные решения. Приведем некоторые оценки для характерных времен в зависимости от прицельного параметра $a$ (скорости $v$, $w$ полагаем $v = w = 0.99$, как и всюду в дальнейшем):

\begin{gather}\label{estimates}
a=0.5  \ \mapsto \ T = -0.44,  \ t' = 0.14,  \ t'' = 28.17,\\
a = 0.05 \ \mapsto \ T = -0.50, \ t' = -0.004, \ t'' = 0.14. \nonumber
\end{gather}
Видно, что в случае $a=0.5$ сигнал о начале взаимодействия приходит после реальной встречи частиц, т.е. частицы начинают вращаться после столкновения (выходят на первый этап в обозначениях работы [6]). В обоих случаях выход на второй этап (согласованное вращение, с момента $t''$) осуществляется после столкновения, в первом случае ($a=0.5$), это происходит существенно позже момента $t=0$. Приведем также оценки для времени прихода сигнала от точки встречи (точки минимального расстояния, соответствующей $t=0$):
\begin{equation}
t_1^0 = \frac{a}{\sqrt{1-v^2}}, \ t_2^0 = \frac{a}{\sqrt{1-w^2}}, 
\end{equation}
\begin{gather}
a=0.5 \ \mapsto \ t_1^0 = 2.51,  \\
a=0.05 \ \mapsto \ t_1^0 = 0.25. \nonumber
\end{gather}
Таким образом, картина столкновения при, например, $a=0.5$ имеет следующий вид: частицы, имея постоянные векторы в изотопическом пространстве, сталкиваются, только после этого в момент времени $\sim~ t_1' = 0.14$ заряды начинают вращаться относительно постоянных векторов (первый этап). Затем приходит сигнал о их столкновении (минимальном расстоянии) в момент времени $\sim t_1^0 = 2.51$, что дает максимальную амплитуду $\widetilde{E_{NA}} \sim d \widetilde{C} / dt$, после чего уже через большой промежуток времени, когда сигнал о вращении гаснет, заряды выходят на второй этап. Однако в случае $a = 0.5$ выход на второй этап происходит в момент времени $t \sim 28$ Фс, что не представляет интереса, так как к этому времени частицы успевают разлететься на достаточно большое расстояние.\\

Уравнение совместности (\ref{condition}) для двух частиц, несущих цветовые заряды $\widetilde{P}$ и $\widetilde{Q}$ с заданным прицельным расстоянием $a$ запишется в виде (везде далее модули цветовых зарядов измеряем в единицах $g$, вынося константу связи в качестве множителя): \\
\begin{gather}\label{rot}
\dot {\widetilde{P}} = \alpha_g \frac{1+vw}{\sqrt{a^2 + (t(v+w) - wt^{**}_{12})^2} + sgn(t) \ w \ \sqrt{(t^{**}_{12})^2 - a^2}}  \ \widetilde{Q}(t-t^{**}_{12}) \times \widetilde{P}, \\
\dot {\widetilde{Q}} = \alpha_g \frac{1+vw}{\sqrt{a^2 + ((v+w) - vt^{*}_{21})^2} + sgn(t) \ v \ \sqrt{(t^*_{21})^2 - a^2}}  \ \widetilde{P}(t-t^{*}_{21}) \times \widetilde{Q}. \nonumber
\end{gather}\\
где времена запаздывания определяются из решения уравнения (\ref{delay}).
Данный вид уравнений не позволяет аналитически (хотя и в три этапа с сшивкой результатов) выписать выражения для положения зарядов в изотопическом пространстве, так как выражение, стоящее перед векторным произведением, является сложной функцией $t$. Более того, в случае не стремящегося к нулю прицельного расстояния на втором этапе ($\sim t''$) уже нельзя пренебрегать запаздыванием, что значительно усложняет получение аналитического ответа. Численное исследование задачи вследствие наличия запаздывания требует неоправданно больших затрат -- запаздывающие конфигурации приходится аппроксимировать тем или иным способом, что в случае малой требуемой точности (и разумного времени выполнения для преследуемых целей) ведет к потере данных, и вращение зарядов учитывается неправильно. В результате этого сигнал с момента встречи частиц, обнаруженный в работе ~\cite{dipodip}, сильно смазывается и становится трудночитаемым. \\

Однако можно сделать некоторые приближения и попытаться привести задачу к виду, пригодному для получения ответа в явной форме. Например, для прицельного параметра $a=0.5$ (дальнейшее исследование будем вести с этим значением $a$) сам процесс столкновения происходит как чисто электродинамический, векторы зарядов сохраняют свое первоначальное положение. Вращение же начинается с момента $t \sim t_1' \sim 0.14$. Можно заметить, что на масштабах $t \sim t_1'$ выполняется неравенство $a^2 (1-v^2) \ll (v+w)^2 t^2$, что позволяет пренебречь прицельным расстоянием как в выражениях для времен запаздывания $t_{12}^{**}, t_{21}^*$, так и в выражении, стоящем перед векторным произведением в (\ref{rot}). 

\subsection{Первый временной этап $t<t'$}
До шкалы времени $t'$ векторы зарядов постоянны: 
\begin{gather*} 
\widetilde{P} = \widetilde{P_T}, \ \ t<t_1', \nonumber \\
\widetilde{Q} = \widetilde{Q_T}, \ \ t<t_2'. 
\end{gather*}
Столкновение происходит при постоянных векторах, что напоминает электродинамическую картину. 

\subsection{Второй временной этап $t' \le t \le t''$}
С учетом описанных выше упрощений уравнения совместности на данном этапе принимают вид (учтено, что $t>0$): 
\begin{gather}\label{rotazero}
\dot {\widetilde{P}} = \alpha_g \frac{1+vw}{t(v+w)}  \ \widetilde{Q_T} \times \widetilde{P}, \\
\dot {\widetilde{Q}} = \alpha_g \frac{1+vw}{t(v+w)}  \ \widetilde{P_T} \times \widetilde{Q}. \nonumber
\end{gather}
С помощью подстановки $\varphi = - ln \  |t|$, уравнения приводятся к виду:
\begin{gather}\label{rotphi}
\widetilde{P'} = \omega \ \widetilde{Q_T} \times \widetilde{P}, \\
\widetilde{Q'} = \omega \ \widetilde{P_T} \times \widetilde{Q}. \nonumber
\end{gather}
Здесь штрих обозначает дифференцирование по переменной $\varphi'$, 
\begin{equation}
\omega = \alpha_g \frac{1+vw}{v+w} \ .
\end{equation}
На данном этапе векторы $\widetilde{P}$ и $\widetilde{Q}$ вращаются вокруг постоянных векторов $\widetilde{P_T}$ и $\widetilde{Q_T}$. 
Для записи решения можно естественным образом выбрать базис для каждой из частиц. С вектором $\widetilde{P}$ связана тройка ортонормированных векторов
\begin{equation}
\widetilde{Q_T} \ , \ \ \widetilde{n}_{P_T} = \frac{\widetilde{P_T} \times \widetilde{Q_T}}{\sin\theta} \ , \ \ \tilde{m}_{P_T} = \widetilde{Q_T} \times \widetilde{n}_{P_T},
\end{equation}
где $\cos\theta = (\widetilde{P_T} \widetilde{Q_T})$.
Аналогично для вектора $\widetilde{Q}$ выбираем
\begin{equation}
\widetilde{P_T} \ , \ \ \widetilde{n}_{Q_T} = - \widetilde{n}_{P_T} \ , \ \ \tilde{m}_{P_T} = \widetilde{P_T} \times \widetilde{n}_{Q_T}.
\end{equation}
Решение может быть представлено в форме
\begin{gather}
\widetilde{P} = \cos \theta \widetilde{Q_T} + \sin \theta \{ \cos \left[ \omega(\varphi - \varphi_1') \right] \tilde{m}_{P_T} - \sin  \left[  \omega (\varphi - \varphi_1') \right] \tilde{n}_{P_T} \} , \\
\widetilde{Q} = \cos \theta \widetilde{P_T} + \sin \theta \{ \cos \left[ \omega(\varphi - \varphi_1') \right] \tilde{m}_{Q_T} - \sin  \left[ \omega (\varphi - \varphi_1') \right] \tilde{n}_{Q_T} \}, \nonumber 
\end{gather}
где использованы обозначения $\varphi_1' = - ln \  t_1', \varphi_2' = -ln \ t_2'$.

\begin{figure}[h!]
\begin{minipage}[h!]{0.59\linewidth}
\center{\includegraphics[scale = 0.20]{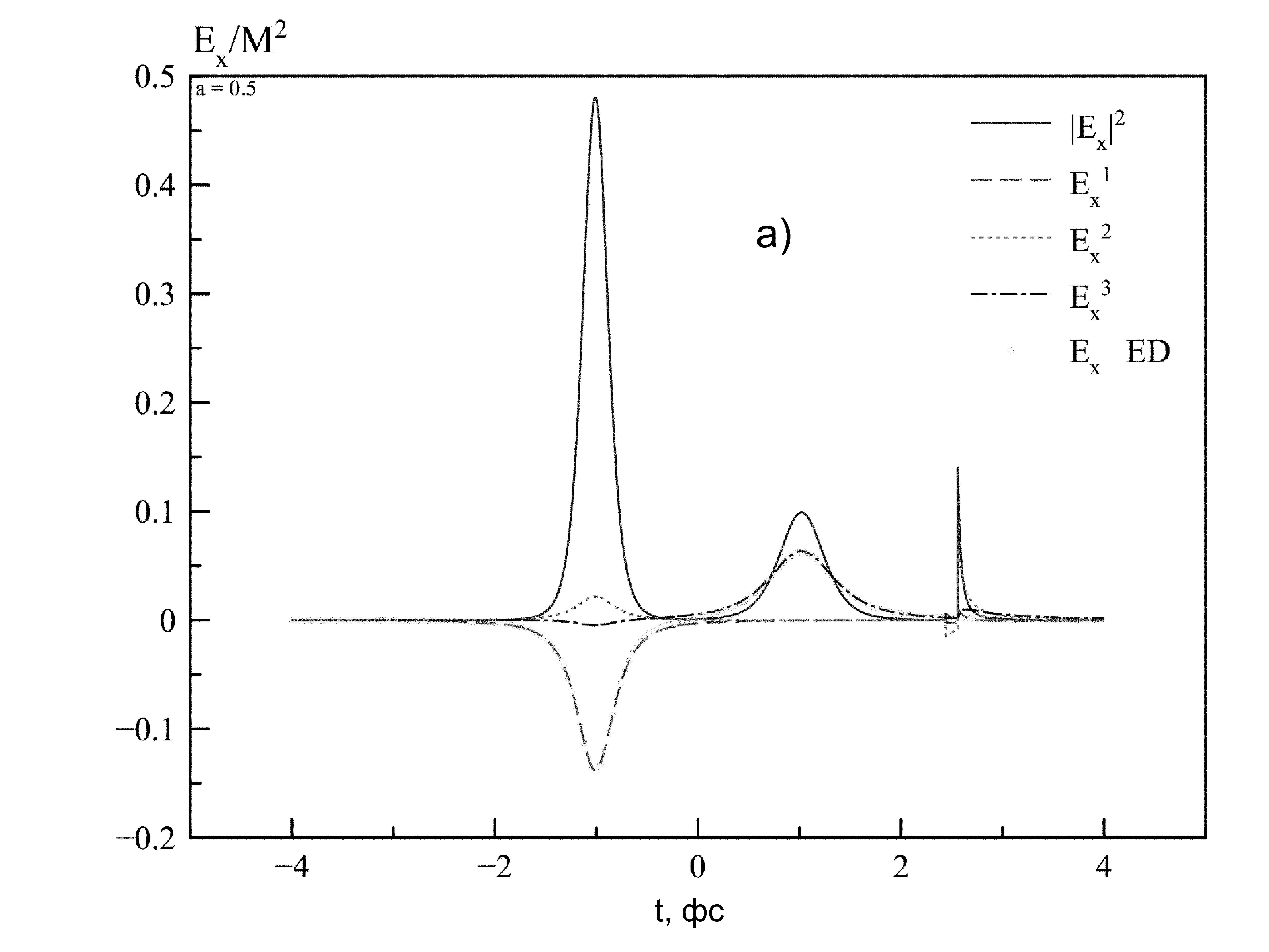}}
\end{minipage}
\vfill
\begin{minipage}[h!]{0.59\linewidth}
\center{\includegraphics[scale=0.20]{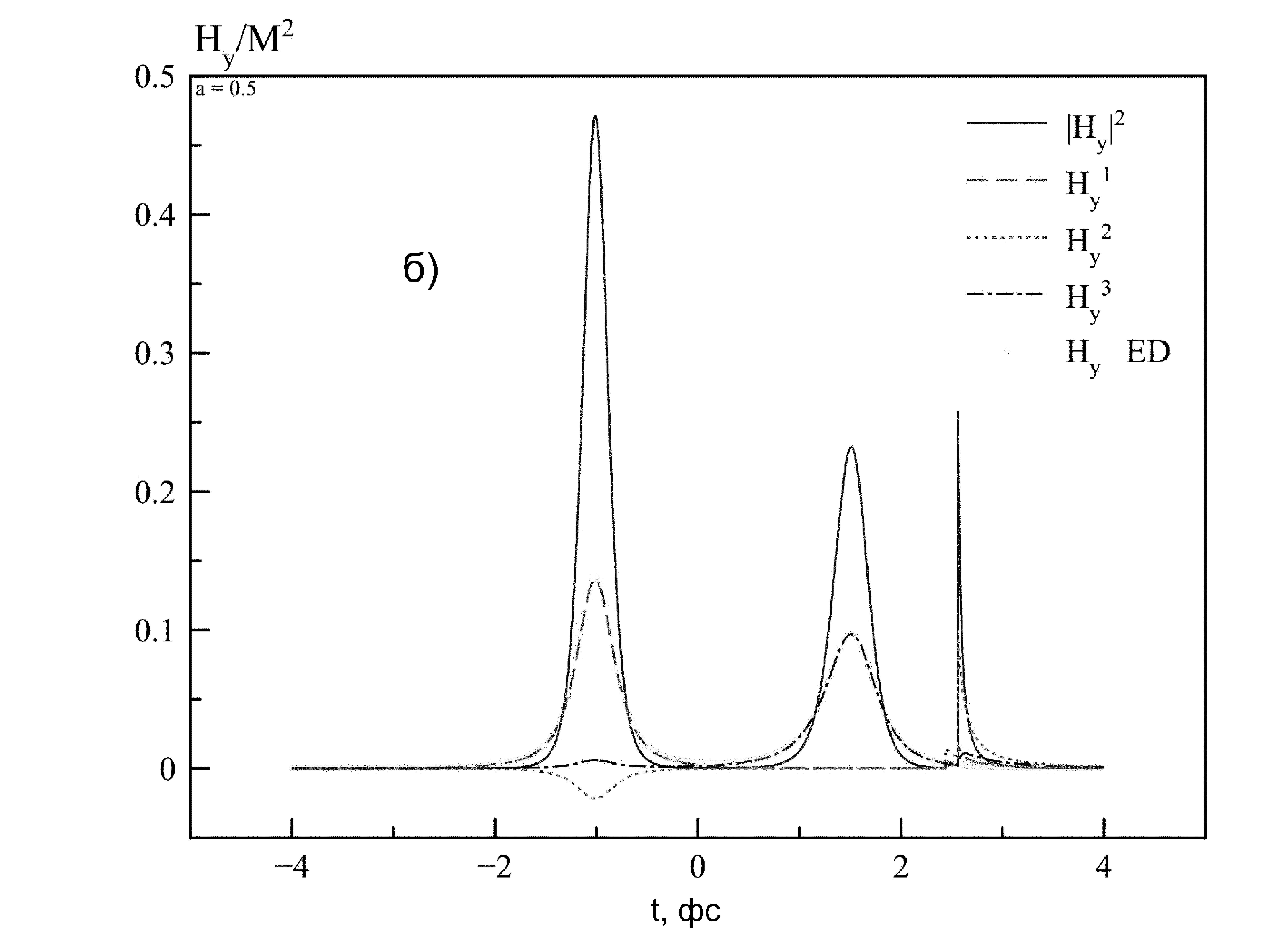}}
\end{minipage}
\caption{Напряженности полей, генерируемых в результате рассеяния частицы на частице, а) - компоненты хромоэлектричсекого поля, б) - компоненты хромомагнитного поля.}
\label{ch-ch}
\end{figure}

На рис. \ref{ch-ch} представлены соответствующие напряженности полей, генерируемых в ходе столкновения в точке наблюдения с координатами $x = 2$ Фм, $z = 1$ Фм при значении прицельного параметра $a = 0.5$, скорости частиц $v = 0.98$, $w = 0.99$. Векторы зарядов частиц единичные и задаются с помощью направляющих углов в изотопическом пространстве. Для первой частицы имеем $\phi_1 = \pi/20, \theta_1 = 0$, для второй - $\phi_2 = -\pi/20, \theta_2 = -\pi/1.95$, $\widetilde{P} = \left( \cos\phi_1 \sin\theta_1, \sin\phi_1 \sin\theta_1, \cos\theta_1 \right)$, $\widetilde{Q} = \left( \cos\phi_2 \sin\theta_2, \sin\phi_2 \sin\theta_2, \cos\theta_2 \right)$ соответственно. 
Напряженности нормированы на $M_\pi ^2$, где $M_pi$ - масса пи-мезона. 
Для удобства сравнения с электродинамической картиной заряды соответствующих электродинамических частиц измеряются в единицах $g$.

Отметим главное отличие полученных результатов от аналогичных, полученных без учета прицельного параметра. Сигнал от вращения частиц (цветовое свечение зарядов) присутствует, но его амплитуда на порядки подавлена по сравнению с ~\cite{dipodip}. Значения напряженностей полей от свечения оказываются даже меньше типичных ``электродинамических'' (кулоноподобных). Указанная особенность объясняется ненулевым значением прицельного параметра (а значит, небесконечной частотой вращения зарядов), а также тем фактом, что вращение зарядов начинается после столкновения.  
\section{Поле цветового заряда и цветового диполя}

\begin{figure}[h!]
\begin{minipage}[h!]{0.59\linewidth}
\center{\includegraphics[scale = 0.20]{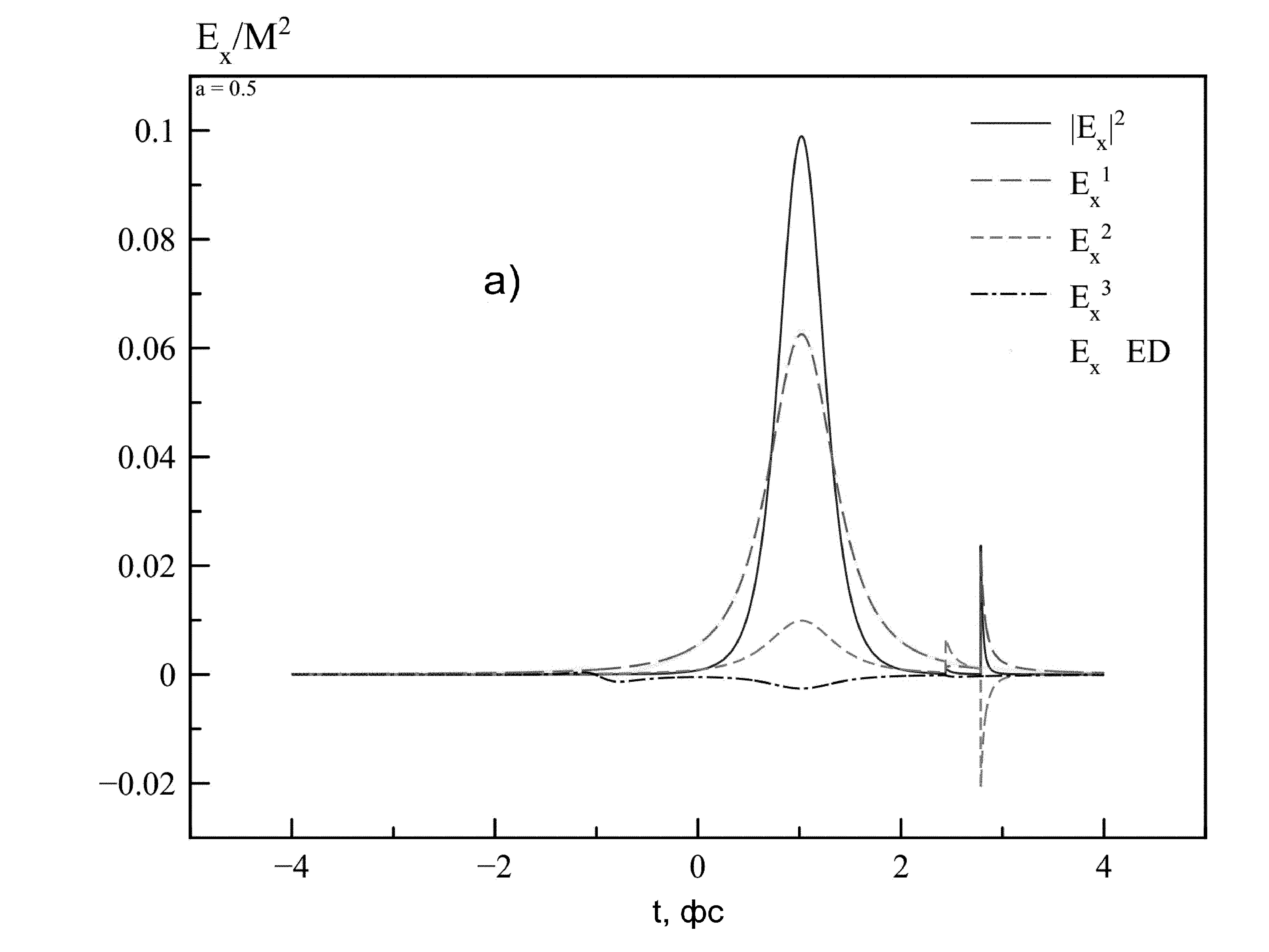}}
\end{minipage}
\vfill
\begin{minipage}[h!]{0.59\linewidth}
\center{\includegraphics[scale=0.20]{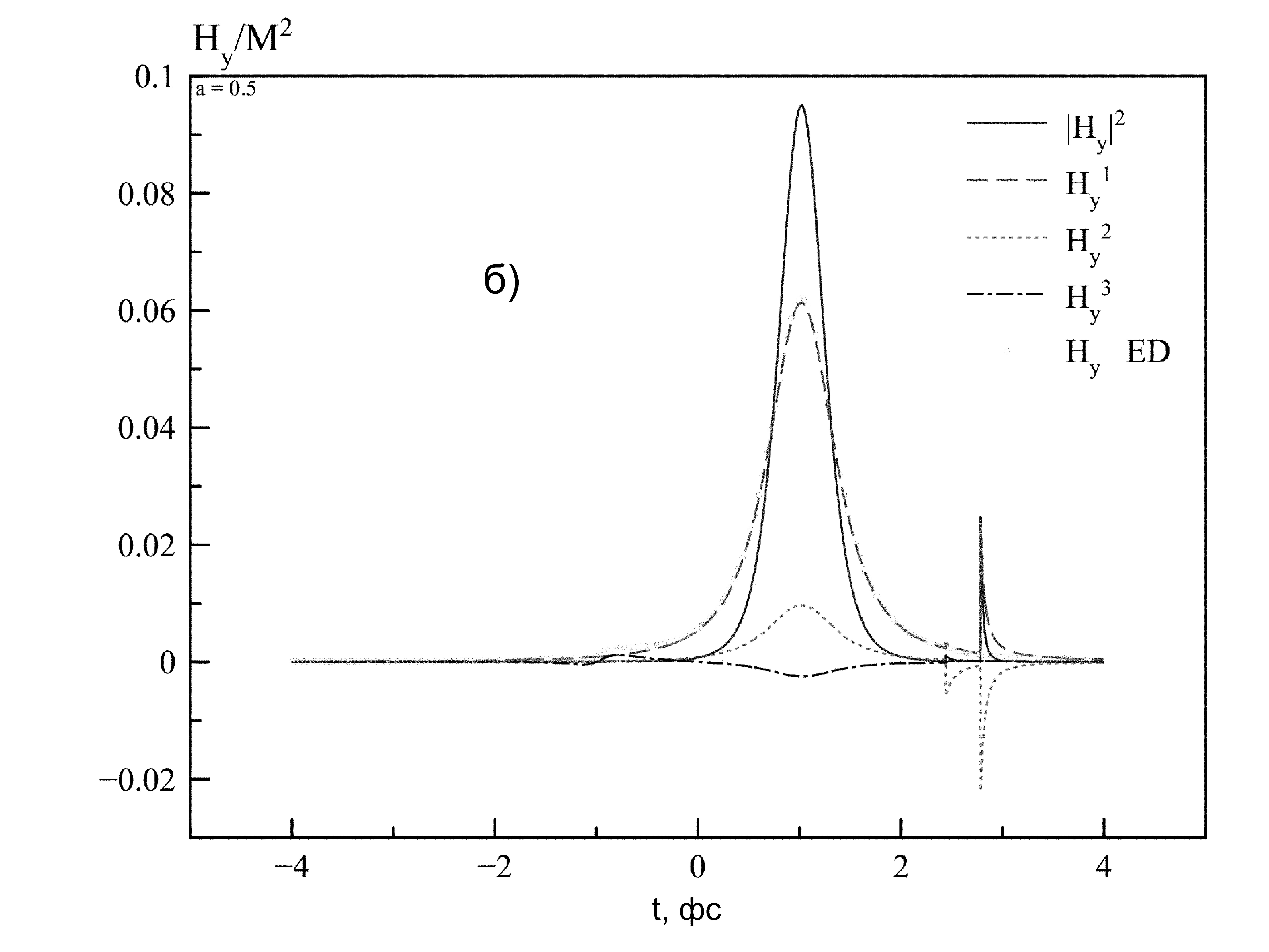}}
\end{minipage}
\caption{Напряженности полей, генерируемых в результате рассеяния частицы на диполе, а) - компоненты хромоэлектричсекого поля, б) - компоненты хромомагнитного поля.}
\label{ch-dip}
\end{figure}

Рассмотрим диполь с расстоянием $\delta$ между зарядами в системе отсчета покоя, движущийся со скоростью $w$ навстречу первой частице с прицельным расстоянием $a$. Заряды диполя обозначим как $\widetilde{Q_2}, \widetilde{Q_3}$. Будем считать диполь ориентированным вдоль направления движения.  В лабораторной системе координат произойдет сильное лоренцево сокращение расстояния $\delta$: $\delta' = (1-w^2) \delta$.
Момент столкновения первой и третьей частиц (под третьей частицей имеем в виду $\widetilde{Q_3}$) $t_3 = \delta' / (v+w)$.
Для рассматриваемых прицельных расстояний приход сигналов к третьей частице описывается аналогично (\ref{estimates}) со сдвигом вправо на величину $t_3$. Вращение зарядов начинается после столкновения с обеими частицами диполя. Соответственно уравнения совместности решаются аналогично уравнениям для двух цветовых зарядов. 
Полное выражение системы уравнений в силу громоздкости приводить не будем. Для упрощения построений можно заметить, что слагаемые, описывающие вращения векторов $\widetilde{Q_2}$, $\widetilde{Q_3}$ друг относительно друга, содержат в числителе множитель $(1-w^2)$, вследствие чего ими можно пренебречь и учитывать вращение зарядов диполя только относительно заряда $\widetilde{P}$.
Результаты численного исследования задачи представлены на рис. \ref{ch-dip}. Значения скоростей и углов зарядов в изотопичеcком пространстве аналогичны рассмотренным в случае столкновения двух цветовых зарядов. При этом в диполе заряды-компаньоны имеют противоположный знак.

\section{Поле двух цветовых диполей}

\begin{figure}[h!]
\begin{minipage}[h!]{0.59\linewidth}
\center{\includegraphics[scale = 0.20]{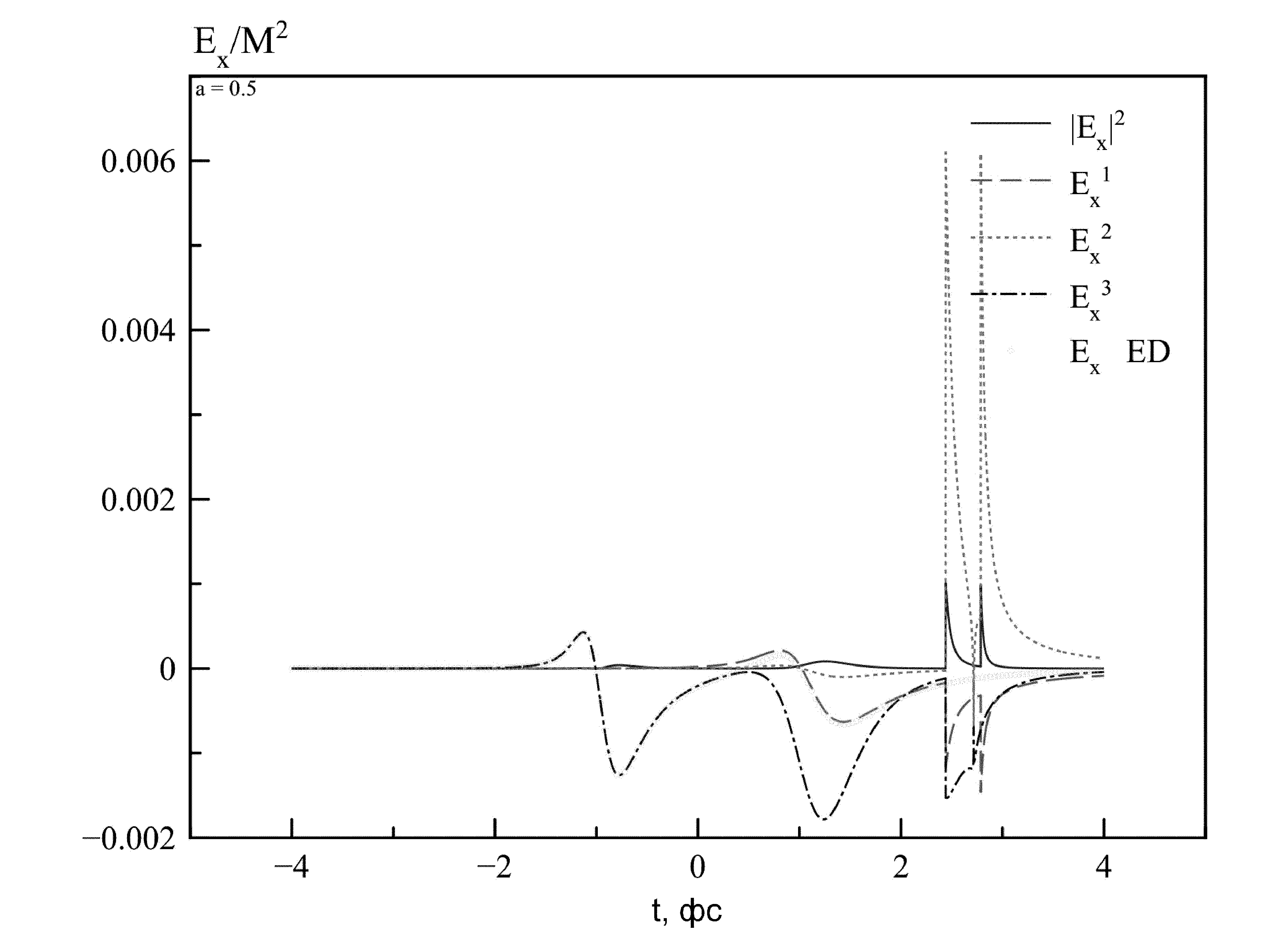}}
\end{minipage}
\vfill
\caption{Напряженности полей, генерируемых в результате рассеяния диполя на диполе.}
\label{dip-dip}
\end{figure}

Для двух цветовых диполей с зарядами $\widetilde{P_1}$, $\widetilde{P_2}$, $\widetilde{Q_1}$, $\widetilde{Q_2}$, движущихся со скоростями $v$, $w$ навстречу друг другу и ориентированных по направлению движения, построение решений осуществляется аналогичным образом. Вращение зарядов в каждом из диполей рассматривается только относительно зарядов из другого диполя. Полное выражение не приводим в силу громоздкости.
Численное исследование представлено на рис. \ref{dip-dip}. Значения скоростей диполей и углов в изотопическом пространстве для зарядов аналогичны рассмотренным выше. Заряды-компаньоны в диполях имеют противоположный знак. 
В случае диполь-дипольного столкновения вклад от сияния превосходит ``электродинамический'', что объясняется компенсацией ``электродинамических'' вкладов от зарядов-компаньонов диполей.

\section{Некоторые возможные физические проявления}

Вообще говоря, из выражения (\ref{EHNA}) видно, что новый член, зависящий от $\widetilde{D}$ = $d\widetilde{C}/dt$, убывает с расстоянием как $1/R$, перенося сигнал от столкновения на большие расстояния. Из-за зависимости типа $1/R$ он может рассматриваться в качестве возможного источника излучения. В этом разделе мы проанализируем этот вопрос. \\

Нетрудно заметить, что по пространственной структуре он мало отличается от ``электродинамического'' (первого слагаемого в выражении (\ref{EHNA}). А значит, сам по себе он излучать не может. Действительно, рассмотрим этот член (далее будем называть его $R$-членом, в силу его зависимости типа $1/R$), порождаемый одной частицей (для определенности положим $\widetilde{C} = \widetilde{P}$). Вычислим поток энергии через поверхность (учитывая, что $\mathbf{n}^2 = 1$):
\begin{gather}
\mathbf{S} = \widetilde { \mathbf { E } }  \times \widetilde { \mathbf { H } }  = \widetilde { \mathbf { E } } \times \mathbf{n} \times \widetilde { \mathbf { E } } = \mathbf{n} \widetilde {E} ^2  - \widetilde { \mathbf { E } } (\widetilde { \mathbf { E } } \mathbf { n} ),   \\ 
(\mathbf{S}\mathbf{n}) = \widetilde { E } ^2 - (\widetilde { \mathbf { E } } \mathbf {n} )^2, \nonumber \\ 
\widetilde { E } ^2 =  \left[ \frac{\widetilde { D } ^2}{R^2} \frac{1 - 2\mathbf {v} \mathbf {n} + v^2 }{(1-\mathbf {v} \mathbf {n})^2} \right] _{t'} , \nonumber \\
(\widetilde { \mathbf  { E } } \mathbf {n} )^2 = \left[ \frac{\widetilde { D } ^2}{R^2} \frac{(1 - \mathbf {v} \mathbf {n})^2 }{(1-\mathbf {v} \mathbf {n})^2} \right]_{t'}, \nonumber \\
\widetilde { E } ^2  = (\widetilde { \mathbf { E } } \mathbf {n} )^2 \ ==> \ (\mathbf{S} \mathbf{n}) = 0. \nonumber
\end{gather}
Таким образом, дополнительный член, связанный с вращением одной частицы, не дает вклада в поток энергии. Учет перекрестных членов  вида $\mathbf{S}^{12} = \widetilde{\mathbf{E}}^{(1)} \times \mathbf{n}^{(2)} \times \widetilde{\mathbf{E}}^{(2)}$ нетривиален, так как для этого необходимо решать задачу рассеяния частиц, но, по всей видимости, они также должны давать нулевой вклад в поток энергии в силу закона сохранения. 
Слагаемые, содержащие произведение $\widetilde{C} \widetilde{D}$, обладают ``неправильным'' знаменателем, в результате чего на больших расстояниях их можно не учитывать. \\
Полученный результат не является необычным, так как вращение зарядов происходит в изотопическом пространстве и не требует затрат энергии. Однако $R$-член, являясь большим по амплитуде и убывая как $1/R$, может переносить информацию о столкновении на большие расстояния (в масштабах сильного взаимодействия), и заключенная в поле энергия может расходоваться на рождение кварк-антикварковых пар. Также он может оказывать влияние на ускорение частиц, причем более значительное по сравнению с электродинамическим (так как не подавлен в продольном направлении множителем $(1-v^2)$, однако нужно не забывать, что он -- следующего порядка малости по $g$), но в силу вращения зарядов является переменным и эффективно может не учитываться. \\

Помимо возможного излучения $R$-член может изменять функцию распределения глюонов (Вайцзеккера -- Вильямса). Интересно сравнить функцию распределения, полученную в классическом приближении в порядке $g^3$ с аналогичной из модели CGC (по аналогии с ~\cite{dipoles}). В ~\cite{dipoles} были получены формулы: 
\begin{gather}\label{weizsacker}
\frac{dE}{dt} = \int \frac{d^2 \mathbf{k_\perp}}{(2\pi)^2} \frac{2g^4}{3m^2} E^2_\parallel ({k_\perp})  \ , \\
E_\parallel ({k_\perp}) = \int d \mathbf{x_\perp} e^{-i (\mathbf{k_\perp} \mathbf{x_\perp})} \langle E_\parallel \rangle \ , 
\end{gather}
из которых можно было получить функцию распределения глюонов
\begin{equation}
f({k_\perp}) = \frac{2g^4}{3m^2} \frac{4\pi g}{dk_\perp} \left[ 1 - \sqrt{\frac{2}{\pi}}(dk_\perp)^{1/2} K_{1/2} (dk_\perp) \right] \ .
\end{equation}
Здесь прицельный параметр $a$ обозначен как $x_\perp$, усреднение ведется по времени столкновения частицы с диполем.
Качественно функция совпадает с аналогичной из ~\cite{struct}, однако в отличие от ~\cite{struct} не имеет инфракрасной расходимости.
$R$-член, не отличающийся по своей структуре от электродинамического (а для двух частиц зависимость $1/R$ превращается в $1/R^2$, так как дополнительный множитель $1/R$ приходит из закона вращения зарядов), не вносит существенного изменения в интеграл (\ref{weizsacker}). Более того, как мы видели, столкновение с диполем на временных масштабах при прицельных расстояниях $a \sim 0.5, 0.05 $ происходит вовсе без вращения зарядов. Учет вращения становится существенным при малых прицельных расстояниях, но одновременно происходит выход за рамки эйконального приближения, и формула (\ref{weizsacker}) становится неприменимой. 

Также стоит отметить, что если при столкновении классический радиус частицы $a = g^2/m$ меньше обратной частоты вращения ($\omega \sim 1/a$, где $\omega$ -- частота вращения, $a$ -- прицельное расстояния), то частица реагирует на усредненное по времени вращения (за него в качестве минимально возможного можно взять время первого этапа) поле. Усреднение такого поля связано с усреднением слагаемых типа скалярных произведений $(\widetilde{P}\widetilde{Q})$ и $\dot{\widetilde{P}} \widetilde{Q}$. В первом случае в $(\widetilde{P}\widetilde{Q})$, несмотря на вращение, всегда имеется постоянная компонента (в случае первого этапа вращения равная начальному углу между векторами зарядов), которая никогда не зануляется при усреднении. Во втором случае (вклад от члена $\dot{\widetilde{P}} \widetilde{Q}$, обусловленного нелинейностью) постоянной компоненты нет, и при усреднении стоит ожидать зануление вкладов данного типа. В частности, для кварков имеем такую ситуацию. 

\section{Цветовое эхо}

Теперь немного подробнее исследуем появление цветового эха, являющегося некоторым проявлением коллективной динамики системы. Именно, зададимся целью выяснить, возможна ли генерация ``сияния'' вследствие эха. \\

Прежде всего, заметим, что в случае $a = 0.5$, $a = 0.05$ эхо для системы типа ``частица + диполь'' не появляется вовсе, так как заряды начинают вращаться после столкновения. Только начиная с $a \sim 10^{-2}$ эффекты типа эха могут проявляться. 
Аккуратное изучение эффекта является сложной вычислительной процедурой, однако, сделав некоторые упрощения, можно получить результат, который, как кажется, будет характерным и более сложных случаях.\\

Для анализа вновь воспользуемся выражением (\ref{rot}). Будем исследовать вращение третьей частицы $\widetilde{Q_3}$ из диполя под влиянием сигнала, пришедшего от частицы с зарядом $\widetilde{P}$ с момента времени $t \to 0$, когда в случае малого $a$ и происходит максимально быстрое вращение $\widetilde{P}$.
Из (\ref{rot}) видно, что в окрестности точки встречи частиц ($t \approx 0$) можно упростить уравнения и привести их к виду, который позволяет написать явное аналитическое решение в удобной для дальнейшего исследования форме: \\
\begin{gather}\label{rotanonzero}
\dot {\widetilde{P} }= \alpha_g  \frac{(1+vw)}{a \ \sqrt{1-w^2}} \widetilde{Q}(t-t^{**}_{12}) \times \widetilde{P},  \\
\dot {\widetilde{Q}} = \alpha_g  \frac{(1+vw)}{a \ \sqrt{1-v^2}} \widetilde{P}(t-t^{*}_{21}) \times \widetilde{Q}. \nonumber
\end{gather}\\
Пренебрегая запаздыванием (если $a$ достаточно мало, оно сказывается только на фазе вращения) и разницей в скоростях $v$ и $w$, можно получить явные выражения для зарядов
\begin{gather}
\widetilde{P} = c \widetilde{\Omega} + s \{ \cos [\omega(t-t'')] \widetilde{m_{p^{''}}} - \sin [\omega(t-t'')] \widetilde{n_{p^{''}}} \},   \\
\widetilde{Q} =c \widetilde{\Omega} - s \{ \cos [\omega(t-t'')] \widetilde{m_{p^{''}}} - \sin [\omega(t-t'')] \widetilde{n_{p^{''}}} \}, \nonumber 
\end{gather}
где 
\begin{equation}
\omega = \alpha_g \frac{(1+vw)}{a \ \sqrt{1-w^2}} \approx \alpha_g \ \frac{(1+vw)}{a \ \sqrt{1-v^2}}  \\, c = \cos \langle \frac{(\widetilde{P^{''}} \widetilde{Q^{''}})}{2} \rangle, s = \sin \langle \frac{(\widetilde{P^{''}} \widetilde{Q^{''}})}{2} \rangle, 
\end{equation}
а опорные векторы определяются как 
\begin{equation}
\widetilde{\Omega} = \frac{\widetilde{P^{''}} + \widetilde{Q^{''}}}{|\widetilde{P^{''}} + \widetilde{Q^{''}}|} \ , \widetilde{n_{p^{''}}} = \frac{\widetilde{P^{''}} \times \widetilde{\Omega}}{|\widetilde{P^{''}} \times \widetilde{\Omega}|} \ , \widetilde{m_{p^{''}}} = \widetilde{\Omega} \times \widetilde{n_{p^{''}}} \ , 
\end{equation}
$\widetilde{P^{''}}, \widetilde{Q^{''}}$ - векторы $\widetilde{P}$ и $\widetilde{Q}$ в момент времени $t^{''}$ (в случае малых $a$ частицы успевают выйти на этап $t \sim t''$ до столкновения).
Стоит отметить, что введение ненулевого (хотя и стремящегося к нулю) прицельного параметра является естественной регуляризацией знаменателя и позволяет проводить аналитические расчеты без разрывов решения. (\ref{rotanonzero}) 
Сигнал от вращения с огромной частотой вектора $\widetilde{P}$ со шкалы времени $t''$ приходит до встречи частиц $\widetilde{P}$ и $\widetilde{Q_3}$. При этом вращение вектора $\widetilde{Q_3}$ приближенно описывается уравнением
\begin{equation}\label{rotthree}
\dot{\widetilde{Q_3}} = \frac{\omega_0}{t_3} \widetilde{P} \times \widetilde{Q_3} \ , 
\end{equation}
где 
\begin{equation}
\omega_0 = \alpha_g \frac{1 + vw}{v+w} \ , 
\end{equation}
$t_3$ -- время встречи первой и третьей частиц, а значение $\widetilde{P}$ берется со шкалы $t''$.\\

Таким образом, мы имеем векторное дифференциальное уравнение вида
\begin{equation}
\dot{\widetilde{Q_3}} (t) = A(t) \widetilde{Q_3} (t) \ , 
\end{equation}
где $\widetilde{Q_3} = \left (Q_3^1, Q_3^2, Q_3^3 \right)$, а матрица $A$ имеет вид 
\begin{equation}
A = \frac{\omega}{t_3} \begin{pmatrix}
0 & -P^*_3 & P^*_2 \\
P^*_3 & 0 & -P^*_1 \\
-P^*_2 & P^*_1 & 0
\end{pmatrix} = A_1 + A_2 \cos{\Omega_0 t} + A_3 \sin{\Omega_0 t} \ , 
\end{equation}
где $P^* = \left(c, \  s_1 \cos\Omega_0 t + s_2 \sin \Omega_0 t, \ -s_1 \sin \Omega_o t + s_2 \cos \Omega_o t \right)$,
\begin{equation}
\Omega_0 = \alpha_g \frac{w}{(1 + \sqrt{1-w^2}) \ a} \frac{1+w}{1-v} \ . 
\end{equation}
Введенные обозначения $s_1, s_2$ определены как
\begin{equation*}
s_1 = s \cos \left[ \omega \left(t'' + \frac{\delta'}{1-v}\right) \right] \ , \ \ s_2 = s \left [\sin \omega \left(t'' + \frac{\delta'}{1-v}\right) \right]. 
\end{equation*}
Решение такого уравнения формально можно представить в виде
\begin{equation}
\widetilde{Q_3} (t) = F(t, t_{e_{1}}) \widetilde{Q_3} (t_{e_1}) \ , 
\end{equation}
где оператор перехода дается выражением 
\begin{equation}
F(t, t_{e_1}) = \int_{t_{e_1}} ^ t A(s) F(s, t_{e_1}) ds = \hat{1} + \int_{t_{e_1}} ^ t A(s_1) ds_1 + \int_{t_{e_1}} ^ t ds_1 \int_{t_{e_1}} ^ {s_1} A(s_1) A(s_2) ds_2 + ...  \ , 
\end{equation}
а время $t_{e_1}$ является временем прихода сигнала о вращении заряда $\widetilde{P}$ к заряду $\widetilde{Q_3}$, $t_{e_1} \sim t'$.
В случае, когда $ t - t_{e_1} = n 2\pi / \Omega_0 $, все интегралы с $A_2$, $A_3$ обнуляются и мы получаем, что
\begin{equation}
\widetilde{Q_3} (t) = e^{A_1 (t - t_{e_1})} \widetilde{Q_3} (t_{e_1}) \ . 
\end{equation}
В матричном виде 
\begin{equation}\label{asympth}
\widetilde{Q_3} (t) = \begin{pmatrix}
Q_{01} \\
Q_{02} \cos{\omega_0 c \tilde{t}} - Q_{02} \sin{\omega_0 c \tilde{t}} \\
Q_{03} \cos{\omega_0 c \tilde{t}} + Q_{03} \sin{\omega_0 c \tilde{t}} \\
\end{pmatrix} \ , 
\end{equation}
где $\widetilde{Q_0}$ -- начальные данные к моменту прихода сигнала, $\tilde{t} = t - t_{e_1}$. \\

\begin{figure}[h!]
\begin{minipage}[h!]{0.59\linewidth}
\center{\includegraphics[scale = 0.7]{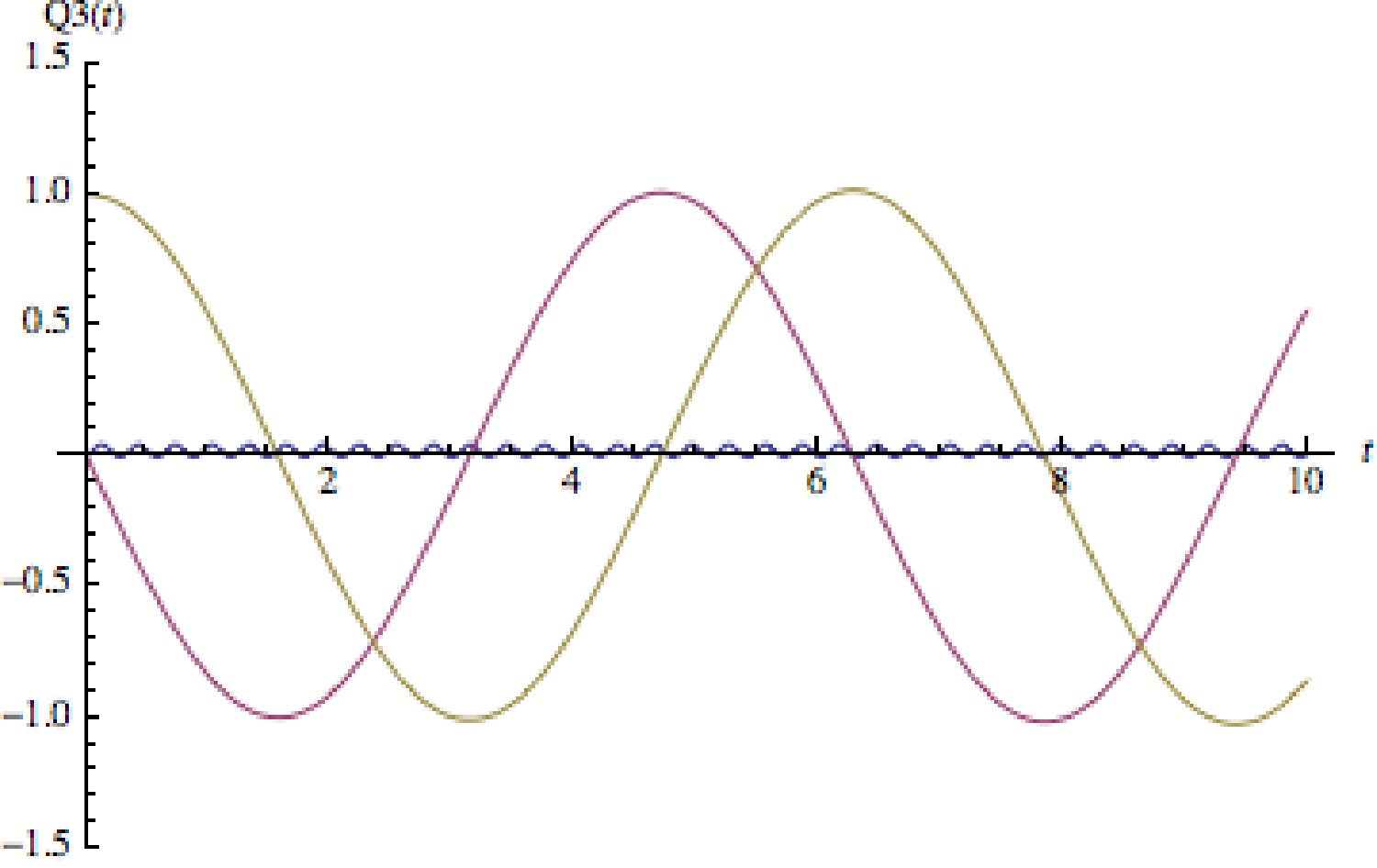} }
\end{minipage}
\vfill
\begin{minipage}[h!]{0.59\linewidth}
\center{\includegraphics[scale=0.7]{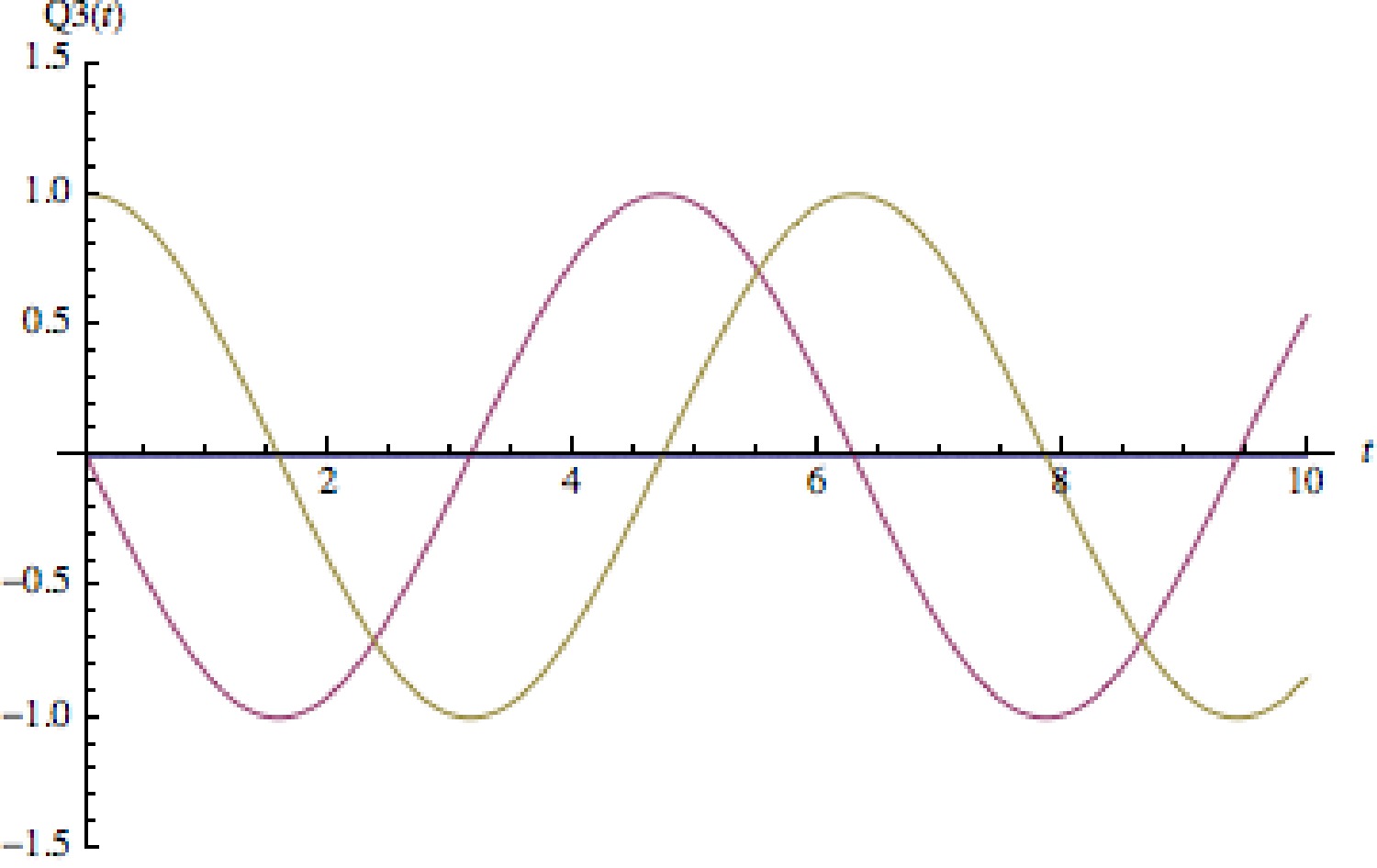} }
\end{minipage}
\caption{Компоненты вектора $Q_3$ для $\omega = 0.1$ (а), $\omega = 10$ (б).}
\label{Echo1}
\end{figure}

В случае $ t - t_{e_1}  = \tilde{t} \ne n 2\pi / \Omega_0 $ интегралы не обнуляются и удается получить лишь асимптотический (при $\Omega_0 \to  \infty$) вид решения. 
Покажем, что решение является ограниченным и в пределе $\Omega_0 \to \infty$ также имеет вид (\ref{asympth}).
При $ \tilde{t} \ne n 2\pi / \Omega_0 $ имеем
\begin{equation}
\Phi (t, 0) = \Phi \left(\tilde{t}, \frac{2\pi}{\Omega_0} \left\lfloor \frac{2\pi \tilde{t} }{\Omega_0} \right\rfloor \right) \Phi \left(\frac{2\pi}{\Omega_0} \left\lfloor \frac{2\pi \tilde{t} }{\Omega_0} \right\rfloor, 0\right) = \Phi \left(\tilde{t}, \frac{2\pi}{\Omega_0} \left\lfloor \frac{2\pi \tilde{t} }{\Omega_0} \right\rfloor \right) \exp{\frac{2\pi}{\Omega_0} \left\lfloor \frac{2\pi \tilde{t}}{\Omega_0} \right\rfloor A_1} \ . 
\end{equation} 
Таким образом, для получения аналитического решения нужно знать матрицу перехода за время, меньшее периода колебаний. Так как получение точного выражения нас не интересует, мы можем качественно описать асимптотическое поведение решений. 
Так как $\|A(\tilde{t}) \| \le \delta$ для некоторого $\delta \ge 0$ при всех $\tilde{t}\geq0$, то
\begin{equation*}
 \| \Phi \left(\tilde{t_2}, \tilde{t_1} \right) \| \leq \exp{\delta(\tilde{t_2} - \tilde{t_1})} \ , \forall \tilde{t_2} \leq \tilde{t_1} \leq 0 \ . 
\end{equation*}
Следовательно,
\begin{gather*}
\| \Phi \left(\tilde{t}, \frac{2\pi}{\Omega_0} \left\lfloor \frac{2\pi \tilde{t} }{\Omega_0} \right\rfloor \right) \| \leq \exp{\left( \frac{2\pi \delta}{\Omega_0} \right)}  , \\
\| \widetilde{Q_3}(t) \| \leq \exp{\left( \frac{2\pi \delta}{\Omega_0} \right)} \| \widetilde{Q}(0) \| \ . 
\end{gather*} 
Также производная является ограниченной при увеличении $\Omega_0$:
\begin{equation*}
\left \| \frac{d\widetilde{Q_3}}{dt} \right \| \leq \| A(t) \| \| \widetilde{Q_3}(t) \| \leq \delta \exp{\left( \frac{2\pi \delta}{\Omega_0} \right)} \| \widetilde{Q}(0) \| \ .
\end{equation*}
Сделанные оценки объясняют вид решения (\ref{asympth}) при $\Omega_0 \to \infty$ и произвольном значении $\tilde{t}$. Соответствующие численные расчеты для значений $\omega = 0.1, \omega = 10$ представлены на рис. \ref{Echo1}. \\

Понятно, что в случае прихода сигнала с более раннего этапа вращение также происходит с конечной частотой. Видно, что несмотря на вращение вектора $\widetilde{P}$ с некоторой (возможно большой) частотой, в целом решение является регулярным с относительно небольшой частотой порядка $\omega_0 c / t_3$, а значит, не может сопровождаться эффектом сияния, который обусловлен огромными частотами вращения (при условии $t_3 \gg a\sqrt{1-v^2}$). Мы рассмотрели только первую итерацию цветового эха, но из рассуждений видно, что дальнейшие итерации также не приведут к эффекту сияния. В рассмотренных в предыдущих разделах ситуации с прицельным параметром порядка $0.5$ Фм, уравнение, аналогичное (\ref{rotthree}) описывает динамику вращения одной из частиц диполя, но на более поздних временных масштабах. Качественно понятно, что и в этом случае ситуация будет похожа на представленную выше.

\section{Заключение}
Мы рассмотрели задачу столкновения элементарных конфигураций частиц с заданным прицельным параметром. Как было показано, в случае разумных значений прицельного параметра столкновения вовсе проходят в ``электродинамическом'' режиме, так как векторы зарядов не вращаются в изотопическом пространстве. Вращение начинается после столкновения, что ведет к появлению нового слагаемого в напряженностях полей. Однако при выбранном значении прицельного параметра $a = 0.5$ Фм вклад ``свечения'' зарядов в напряженности полей оказывается существенно меньше аналогичного, возникающего при лобовом столкновении. При меньших значениях прицельного параметра вклад ``свечения'' растет, однако разумно обрезать дальнейшее уменьшение прицельного параметра при достижении некоторого значения. Показано, что непосредственно на динамику эффект ``свечения'' не оказывает существенного влияния. \\

Исходя из анализа ситуации с цветовым эхо, видно, что сделанный вывод оказывается справедливым и для более сложных конфигураций частиц, так как жесткие глюонные поля могут генерироваться только в результате непосредственного столкновения, а не за счет каких-либо эффектов типа эха. \\

Практический вывод о возможности описания системы в приближении ''чистой'' электродинамики с усиленной константой связи особенно подтверждается при наличии прицельного параметра порядка нескольких десятых долей ферми, как естественного регуляризатора. В случае ультрарелятивистских скоростей и на расстояниях порядка 1 Фм нелинейность, связанная со слагаемыми, пропорциональными $g$, не изменяет кулоноподобного поведения.  \\

Автор выражает благодарность {\fbox {С. В. Молодцову}} и А. М. Снигиреву за замечания и полезные обсуждения.



\clearpage

\clearpage

\begin{center}
\large \bfseries \MakeTextUppercase{
Classical gluon fields of relativistic color charges
}
\end{center}

\begin{center}
\bfseries 
    A. S. Zadora
\end{center}

\begin{center}
\begin{minipage}{\textwidth}
\small
\vspace{1cm}
The aim of present work is to consider in more details recently theoretically observed exotic ``color charge glow'' effect and  possible physical effects related to ensembles of color charged particles on the classical level. We study ways of occurrence of ``color charge glow'' in arbitrary systems of point massive particles which could be split into elementary configurations like color charges and color dipoles. Also the impact of this effect on dynamics of such systems is studying. For the natural regularization of expressions we consider collisions of particles with given impact factor. We show that with reasonable values of impact factor collisions could be quite of ``electrodynamical'' (Coloumb-like) regime and the contribution of ``glow'' turns out to be suppressed in comparison with ``electrodynamical'' picture. It follows from the analysis of ``color echo'' effect that this result is still correct if we consider more complicated configurations of particles because hard gluon fields could arise only due to immediate collisions but not as a result of some effects like ``color echo''.
\end{minipage}
\end{center}

\clearpage

\begin{center}

\begin{enumerate}

\item Напряженности полей, генерируемых в результате рассеяния частицы на частице, а) - компоненты хромоэлектричсекого поля, б) - компоненты хромомагнитного поля. 
\item Напряженности полей, генерируемых в результате рассеяния частицы на диполе, а) - компоненты хромоэлектричсекого поля, б) - компоненты хромомагнитного поля.
\item Напряженности полей, генерируемых в результате рассеяния диполя на диполе.
\item Компоненты вектора $Q_3$ для $\omega = 0.1$ (a) и $\omega = 10$ (б).

\end{enumerate}

\end{center}

\end{document}